\documentclass[twocolumn,dvipsnames,twocolappendix]{aastex63}
\usepackage{amsmath, amssymb, graphics, bm, graphicx,color,ulem}
\usepackage{appendix}
\usepackage{verbatim}

\usepackage{xcolor}

\newcommand{\w}{\color{blue}}
\newcommand{\y}{\color{red}}
\newcommand{\x}{\sout}
\newcommand{\z}{\color{orange}}

\newcommand{\mathsym}[1]{{}}
\newcommand{\unicode}[1]{{}}

\newcommand{\om}{\omega}

\newcommand{\der}{\text{d}}

\newcommand{\Ms}{M_\star}

\newcommand{\Msun}{{\rm M}_\odot}

\newcommand{\Porb}{P_{\rm orb}}

\newcommand{\e}{{\rm e}}

\newcommand{\be}{\begin{equation}}
\newcommand{\ee}{\end{equation}}
\allowdisplaybreaks

\graphicspath{{./}{figures/}}

\begin{document}

\title{A Tale of Two Circularization Periods}

\correspondingauthor{J.~J. Zanazzi}
\email{jzanazzi@cita.utoronto.ca}

\author{J.~J. Zanazzi}
\affiliation{Canadian Institute for Theoretical Astrophysics, University of Toronto, 60 St. George Street, Toronto, ON M5S 3H8, Canada}


\begin{abstract}
We re-analyze the pristine eclipsing binary data from the \textit{Kepler} and TESS missions, focusing on eccentricity measurements at short orbital periods to emperically constrain tidal circularization.  We find an average circularization period of $\sim$6 days, as well as a short circularization period of $\sim$3 days for the \textit{Kepler}/TESS field binaries.  We argue previous spectroscopic binary surveys reported longer circularization periods due to small sample sizes, which were contaminated by an abundance of binaries with circular orbits out to $\sim$10 days, but we re-affirm their data shows a difference between the eccentricity distributions of young ($<$1 Gyr) and old ($>$3 Gyr) binaries.  Our work calls into question the long circularization periods quoted often in the literature.
\end{abstract}

\section{Introduction}

A long-standing puzzle in stellar physics is how tides dissipate in binaries. Just like that in the Earth-Moon system, tides excited by near-by companions can be damped by internal friction, leading to orbital circularization and spin synchronization. This explains why short period binaries tend to be circular, and likely spin synchronized. However, unlike the Earth-Moon system, stars are nearly  ideal fluid with negligible molecular viscosity. And despite multiple decades of theoretical efforts, the origin of this friction remains unclear.

There are two classes of tidal theories. One, championed by Zahn and collaborators, posits that 
tidal friction arises from damping of the equilibrium tidal response in the turbulent convection zones
\citep[e.g.][]{Zahn(1966),Zahn(1977),Zahn(1989),ZahnBouchet(1989)}.  The main uncertainty in these theories is the efficiency of  turbulent damping, especially in the so-called 'fast-tide' regime. Competing theories \citep{GoldreichNicholson(1977),GoodmanOh(1997)} yield  friction estimates that differ by orders of magnitude \citep{Penev(2007),OgilvieLesur(2012),Duguid(2020a),Duguid(2020b),VidalBarker(2020a),VidalBarker(2020b)}.\footnote{Recently, \cite{Terquem(2021a),TerquemMartin(2021)} proposed an un-suppressed source of dissipation from turbulent convection, but see \citet{BarkerAstoul(2021)} for a rebuttal.}

The other class of theories focus on 'dynamical' tidal responses. These consider the dissipation of tidally forced oscillations:  internal gravity-modes damped by radiative diffusion and turbulent convection \citep[e.g.][]{Zahn(1975),Zahn(1977),NorthZahn(2003),GoodmanDickson(1998),Terquem(1998)}, and possibly nonlinear wave-breaking
\citep[e.g.][]{GoodmanDickson(1998),OgilvieLin(2007),BarkerOgilvie(2010),Barker(2020)}; rotationally-supported inertial waves damped by viscosity \cite[e.g.][]{OgilvieLin(2007),Wu(2005b),GoodmanLackner,LinOgilvie(2018),LinOgilvie(2021)}.
However, the generally weak tidal forcing and the transient nature of tidal resonances \citep{Terquem(1998)} may conspire to make these forced oscillations unimportant. This then stimulates the recent development of the so-called `resonance locking' theories \citep{SavonijePapaloizou(1983),SavonijePapaloizou(1984),WitteSavonije(1999b),WitteSavonije(2001),WitteSavonije(2002),SavonijeWitte(2002),FullerLai(2012),Burkart(2012),Fuller(2017),MaFuller(2021)} whereby tidal resonances are prolonged by stellar evolution. Recent work shows resonance locking onto gravity modes efficiently circularizes binaries during the pre-main sequence, with comparatively little additional circularization during the main-sequence \citep{ZanazziWu(2020)}.

\begin{figure}
\includegraphics[width=\columnwidth]{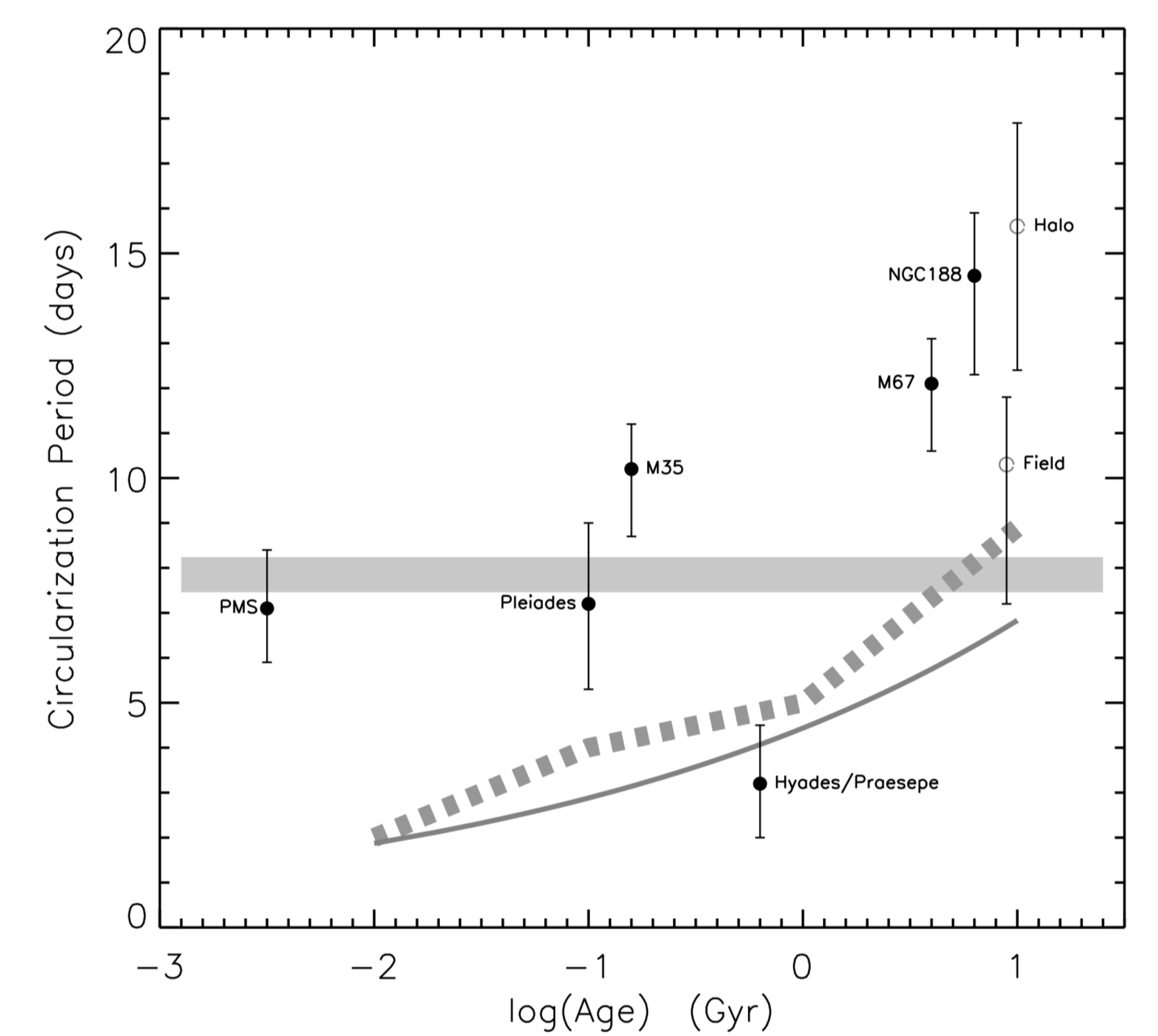}
\caption{Figure 9 from \citet{MeibomMathieu(2005)}, reproduced with permission.  Measurements of binary orbits in different clusters and environments appear to show that the circularization period (see text for definition) rises gradually during the main-sequence,  from about 8 days for the pre-main-sequence populations to about 15 days in the oldest populations. The curves are predictions from  various theories, see that paper for detail. 
}
\label{fig:meibom_dist}
\end{figure}

Interestingly, while theorists are clearly excited by and invested in this problem, there are scant observational constraints. The most notable exception is the series of works by Mathieu and collaborators \citep{Latham+(1992),Mathieu+(2004),MeibomMathieu(2005),Meibom+(2006),Geller+(2012),Geller+(2013),Milliman(2014),Leiner(2015),Nine+(2020),Geller(2021)}. Using a sample of binary orbits collected painstakingly through radial-velocity monitoring, they measured the so-called `circularization period', the period out to which most binaries appear to be circular, for stellar clusters at different ages. Figure \ref{fig:meibom_dist} is the culmination of their body of works \citep[see][for updates]{Nine+(2020)}, where solar-type binaries younger than $\sim 1$ Gyrs are shown to be circular out to about $8$ days, while 
this value rises to $\sim 15$ days for those in the oldest open clusters and the halo. Given that the strength of tidal interactions drops steeply with increasing binary separation, these long circularization periods suggest that internal friction in stars is much higher than most reasonable estimates, and that tidal circularization occurs mostly during the main-sequence. This poses a significant constraint on the tidal theories, and remains an outstanding problem in astrophysics \citep[e.g.][]{Mazeh(2008)}.

Almost a decade after the pioneering work of \citet{MeibomMathieu(2005)}, there has been little progress in examining the robustness of their conclusions, which is the goal of this paper. We are aided by results from a number of recent surveys, such as eclipsing binaries from the {\it Kepler} and TESS photometric missions \citep[e.g.][]{Milliman(2014),VanEylen(2016),Triaud(2017),Windemuth(2019),JustesenAlbrecht(2021)} and radial-velocity binaries from the SDSS spectroscopic survey \citep{PW-goodman,Price-Whelan(2017),Price-Whelan(2020),Kounkel}. By analysing the eclipsing binary data, we find most binaries circularize interior to $\sim$6 days, with a sub-population of binaries circularizing interior to only $\sim$3 days, a conclusion which differs drastically from \cite{MeibomMathieu(2005)}.  In the following, we present the eclipsing binary data (\S \ref{sec:EB_data}), and how we calculate the circularization period (\S \ref{sec:emperical_circ}). In \S\ref{sec:Discuss}, we re-examine the \citet{MeibomMathieu(2005)} data to understand the origin of our discrepancy, followed by a  brief discussion on the theoretical implications  of our results.

\section{Eclipsing Binary Data}
\label{sec:EB_data}

\begin{figure}
\includegraphics[width=\columnwidth]{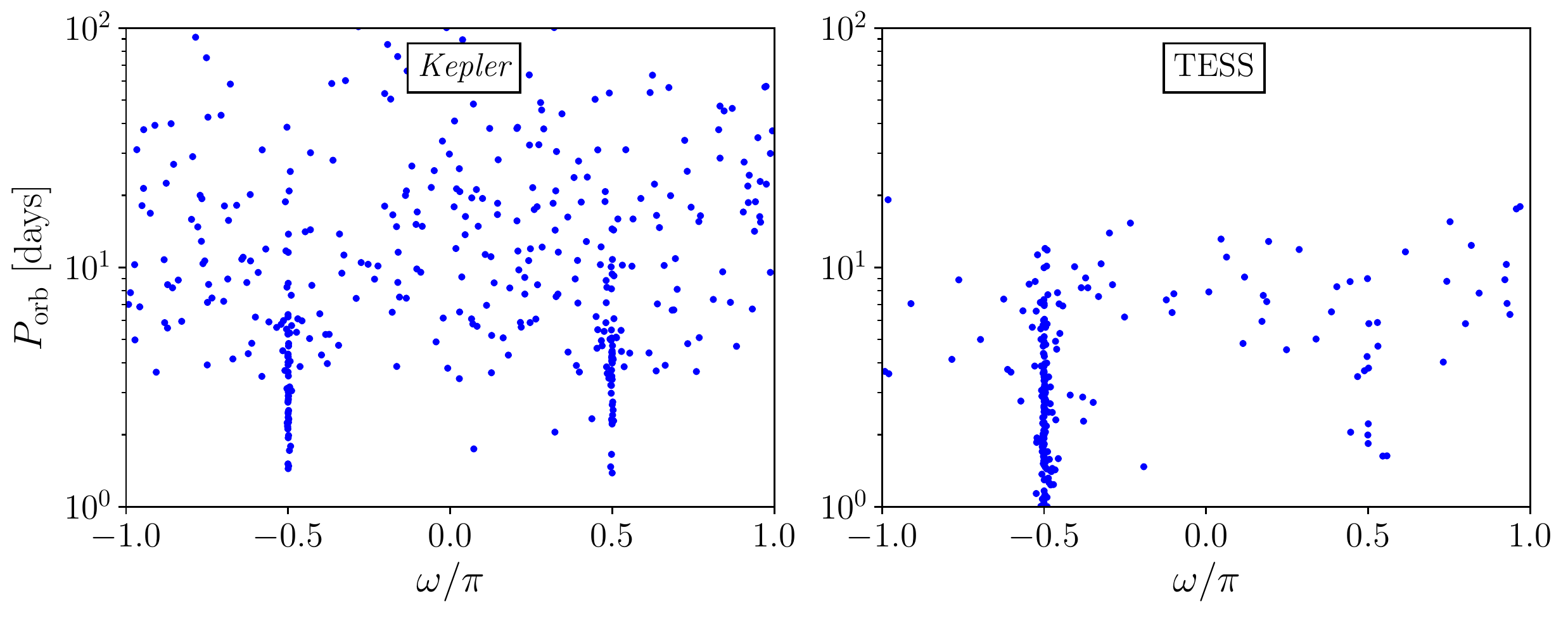}
\caption{Calculated longitude of pericenter $\om$ over orbital periods $P_{\rm orb}$, using the EB data from \cite{Windemuth(2019)} and \cite{JustesenAlbrecht(2021)}.  The clustering around $\om \approx \pm \pi/2$ is not real, but indicates a bias due to large $e\sin\om$ measurement errors, which tends to produce measurements with $|e\sin\om| \gg |e\cos\om|$. We discard $e\sin\om$ measurements in this work.
}
\label{fig:period_pomega}
\end{figure}

We analyze the eclipsing binaries (EBs) discovered by the {\it Kepler} \citep{Borucki+(2010)}  and TESS \citep{Ricker+(2015)} missions.
Lightcurves of EBs reveal the orbital eccentricities when both the primary and the secondary eclipses are detected \citep[for a tutorial, see e.g.][we briefly recap in   Appendix~\ref{app:ecc}]{Winn(2010)}. Qualitatively, while primary and secondary transits in circular orbits occur exactly half an orbital period apart, eccentric orbits do not (unless the eccentricity vectors are fortuitously aligned with the line-of-sight); durations of the two transits also encode information about the eccentricity. Hence, the precise photometric data from {\it Kepler} and TESS can reveal projected eccentricicity values as minute as $\sim 10^{-3}$. 

With nearly continuous photometric monitoring that spans 4 years, {\it Kepler} discovered $\sim 3000$ eclipsing binaries  with  periods reaching out $\sim 3$ yrs \citep{Prsa(2011),Slawson(2011),kirk}. These form a valuable sample for studying tidal circularization. \citet{Windemuth(2019)} re-analyzed {\it Kepler} lightcurves to derive 
orbital parameters ($P_{\rm orb}$, $e\cos\omega$ and $e\sin\omega$, where $\omega$ is the the longitude of pericentre) for $728$ systems. They then inferred stellar effective temperatures, radii, masses using \textit{Gaia} data  and  stellar  isochrones. They also provided estimates for the stellar ages, but caution that these are  likely unreliable. We believe this is indeed the case -- although one expects no more than $\sim$1\% of stars in the \textit{Kepler} sample to have ages younger than $\sim 10^8$ years, $\sim$20\% of the binaries in the \cite{Windemuth(2019)} catalogue do so.  In this work, we discard their age information.
We then remove the 35 systems identified by \citet{Windemuth(2019)} as showing transit timing variations (likely due to a tertiary companion), as well as  48 binaries that are very tight and exhibit large ellipsoidal variations (ones with morphology parameters $ > 0.5$). Both cuts do not impact our study significantly.

Compared to the {\it Kepler} mission, the  TESS mission has a larger field-of-view but a
shorter monitoring duration. \citet{JustesenAlbrecht(2021)} extracted $\sim 1000$ EBs with periods extending up to $20$ days. Other than  the orbital parameters, they 
also inferred stellar parameters (stellar radii and effective temperatures, but not masses) by combining TESS folded light-curves with the binary component Spectral Energy Distributions, the latter obtained by combining broadband photometry with \textit{Gaia} DR2 parallaxes. For our study, we append this sample to the above {\it Kepler} sample.

As we are mostly interested in the tidal dynamics of FGK stars, we retain only systems with primary masses
within the range $0.8 \Msun \le \Ms \le 1.4 \, \Msun$ for the \textit{Kepler} sample, and effective temperatures within $4500 \, {\rm K} \le T_{\rm eff} \le 7000 \, {\rm K}$ for the TESS sample. We are left with a total of 524 EBs with $\Porb$ values between 1 and 100 days. We use these collectively to determine the sample circularization periods. For comparison, \citet{MeibomMathieu(2005)} typically used a few dozen binaries to do so for a given cluster.

In our analysis, we will only use the measured $e\cos\omega$ values, but discard those for $e\sin\omega$. The former are determined from centroiding the primary and secondary transits and so can reach supreme precision. The error margin on $e\cos\omega$ is on the  order of $5\times 10^{-4}$ \citep[Windemuth, private communications,][]{JustesenAlbrecht(2021)}, affording us useful information on the eccentricity over several decades.  In contrast, the values of $e\sin\omega$ are determined by measuring the relative widths of the transits and suffer from much larger uncertainties. For circular orbits, one often measures $|e\sin\omega| \gg |e\cos\omega|$. This is indeed seen in data (Fig. ~\ref{fig:period_pomega}), where the inferred $\omega$ values have an unnatural clustering round $\omega = \pm \pi/2$, a problem also pointed out by \citet{VanEylen(2016),JustesenAlbrecht(2021)}. 

\section{The Circularization Period}
\label{sec:emperical_circ}

\begin{figure*}
\includegraphics[width=\textwidth]{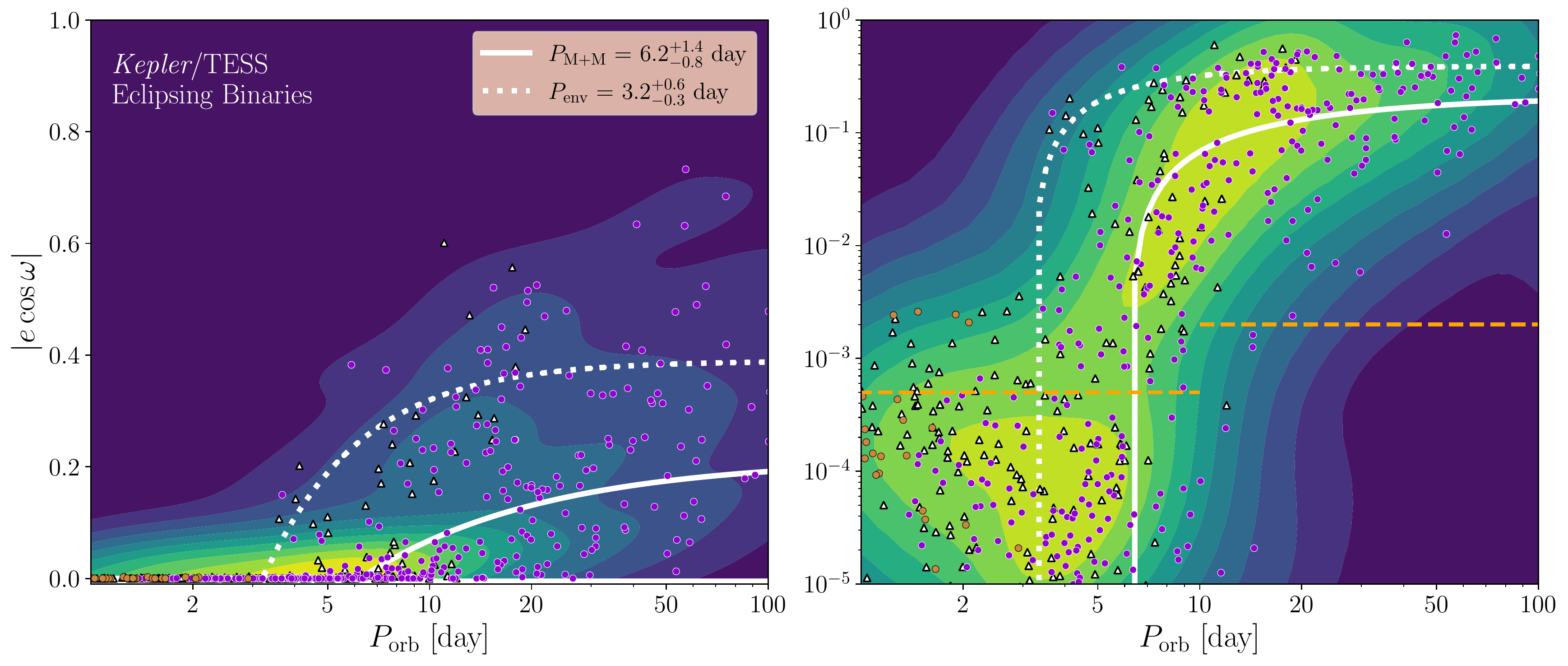}
\caption{Eccentricity data from Kepler (purple points) and TESS (white triangles), with contours denoting smoothed Kernel Density Estimates of the  eccentricity measurements. The dashed orange line denotes typical errors in $|e \cos \om|$ measurements in the \textit{Kepler}/TESS dataset, while light-brown points denote removed binaries with $\texttt{morph}>0.5$ (see text for discussion).  White dotted and dashed lines denote fits for the mean ($P_{\rm M+M}$) and envelope ($P_{\rm env}$) eccentricity distributions.  Other eccentricity distribution parameter fits are constrained to $\alpha_{\rm M+M} = 0.21^{+0.79}_{-0.06}$, $\beta_{\rm M+M} = 0.8^{+1.2}_{-0.7}$, $\alpha_{\rm env} = 0.39^{+0.06}_{-0.05}$, $\beta_{\rm env} = 1.5^{+1.0}_{-0.6}$.
}
\label{fig:data-ecc-period}
\end{figure*}

Here, we use the observed EB sample to measure the so-called circularization period, the orbital period out to which most binaries are tidally circularized.  Figure~\ref{fig:data-ecc-period} displays the EB data in the period-eccentricity (actually
 $|e\cos\omega|$) space. We present them in both linear and logarithmic eccentricities. The latter information is unique to EBs and a testament to the power of transit missions. The eclipsing binaries clearly exhibit an upper envelope in eccentricity that rises with orbital period, a trade-mark of tidal circularization. However, although binaries below $
 \sim$3 days have infinitesimally-small eccentricities (the `waterfall' feature in logarithmic eccentricity, see Fig.~\ref{fig:synth-ecc-period}), there exists a core of circular binaries with periods extending to as long as $\sim 10-20$ days. This latter feature has an unclear origin, but can bias analysis that only include a small number of binaries (see \S\ref{sec:Discuss}).

To quantitatively constrain the properties of the observed EBs, 
we adopt the following functional form for the eccentricities 
\be
e(\Porb) = \left\{
\begin{array}{ll}
0 & \Porb \le P_{\rm circ}\, ,
\\
\alpha \left[1 - \left( \dfrac{P_{\rm circ}}{\Porb} \right)^\beta \right] & \Porb > P_{\rm circ}\, ,
\end{array} \right. 
\label{eq:fe}
\ee
with three free parameters  $P_{\rm circ}$, $\alpha$ and $\beta$. This form differs slightly from that in
\cite{MeibomMathieu(2005)}: $\alpha (1-\exp[0.14(P_{\rm circ}-\Porb)])$. 
Our form is slightly simpler, and 
returns similar values for $P_{\rm circ}$ when applied to the samples used in \citet{MeibomMathieu(2005)}. 

We then follow two different ways to characterize the circularization period. For the first, we follow \citet{MeibomMathieu(2005)} to minimize the metric
\begin{equation}
\chi^2 = \sum_i \left(e(P_i) - |e_i \cos\omega_i|\right)^2\, ,
\label{eq:chi2}
\end{equation}
where the summation is over all binary systems. This is similar to the procedure in  \citet{MeibomMathieu(2005)} (we neglect measurement errors), and we denote the circularization period thus obtained as  $P_{\rm M+M}$. 

Because the EB sample is complex, we find that different metrics can give rise to different values of $P_{\rm circ}$. The original $P_{M+M}$ fits for the mean of the eccentricity-period distribution, but does a poor job of describing the distribution of the most eccentric binaries. So we devise a metric that emphasizes the upper envelope of the eccentricity distribution.  To do this, we first separate binaries into  $20$ logarithmic period bins, and pick the most eccentric binary within the bin.  We then fit equation~\eqref{eq:chi2} to these maximum eccentricities at binned orbital periods.  We denote the best fit $P_{\rm circ}$ thus obtained as $P_{\rm env}$, for the ``envelope'' of the distribution.

Uncertainties on the circularization period are dominated by the finite sizes of our samples, as opposed to measurement errors on $|e\cos\omega|$ (very small), or the strategy of using $|e\cos\omega|$ as a proxy for the full eccentricity (see Fig.~\ref{fig:proj-check}). To estimate these, 
we  estimate the errors on the circularization period via bootstrapping, randomly re-selecting 524 binaries from our sample (so some measurements are counted more than once, while others are left out), and fit the data for the Meibom \& Mathieu $P_{\rm M+M}$ and envelope $P_{\rm env}$ circularization periods.  We repeat this process 3000 times, and calculate the median and $1\sigma$ uncertainties from the distribution of fitted values.

As is clear from Fig.~\ref{fig:data-ecc-period}, the observed population harbour a cold-core of circular binaries that extend well beyond the shortest circularization period $P_{\rm env}$ \citep[also see][]{Triaud(2017),Kjurkchieva(2017)}.
This population severely biases $P_{\rm M+M}$ (sensitive to the average eccentricity) to longer values, compared to that for $P_{\rm env}$ (sensitive to the upper envelope). In the following (\S \ref{sec:Discuss}), we argue that circularization periods determined using a smaller sample are strongly influenced by the presence of this cold core, and it may explain much of the differences in conclusions between our work and that of \citet{MeibomMathieu(2005)}.

\section{Discussion}
\label{sec:Discuss}

In this section, we review previous works that have bearings on our results, the most notable one being that of \citet{MeibomMathieu(2005)}. We then turn to a brief discussion on the implications for the theory of tidal circularization.

\subsection{Comparison to Previous Studies}

Contrary to what we conclude here,  \citet{MeibomMathieu(2005)} found a long circularization period of $\sim$8 days for young binaries, which increases as the stars age.  We have found that a number of factors can partially account for our differences, but some discrepancies persist. 

The first, and likely the most important factor is the difference in our sample sizes. While we use  524 binaries to collectively determine a circularization period, \citet{MeibomMathieu(2005)} determined one such period for each cluster, based on no more than a couple dozen binaries, and the numbers that are most constraining (orbital periods from a few to a few tens of days) are even smaller. To emulate the impacts this cast on the results, we randomly draw $N_b$ binaries from the 524 EB sample, and re-determine the circularization period. We only include EBs with periods shorter than $100$ days, as only they have power to constrain the circularization period.

\begin{figure}
\includegraphics[width=\columnwidth]{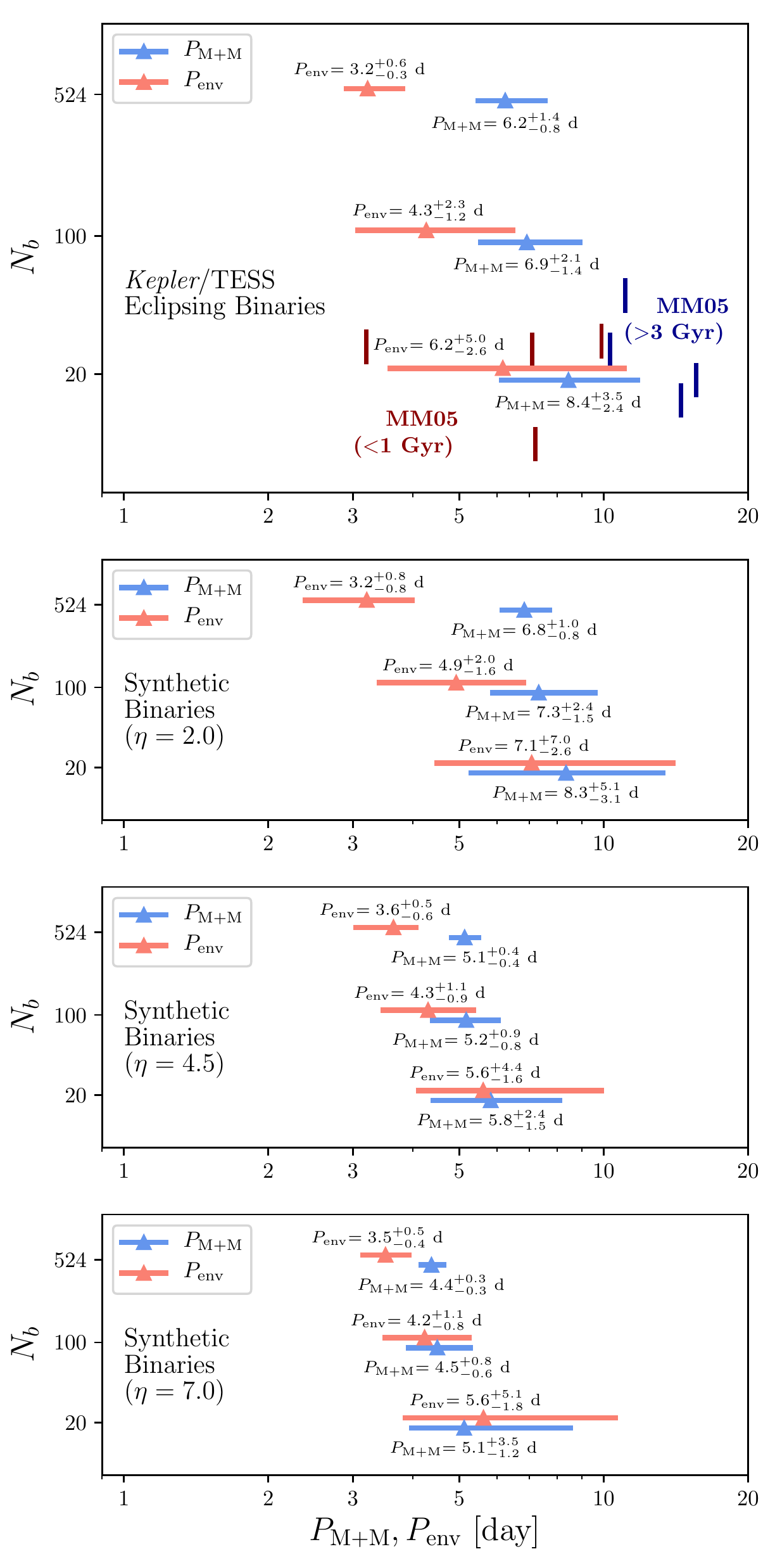}
\caption{\textit{Top panel}: Circularization periods $P_{\rm M+M}$ (blue) and $P_{\rm env}$ (orange) for the \textit{Kepler}/TESS eclipsing binaries, varying the number of binaries $N_b$ drawn from the data-set, with triangles and lines denoting the median and 1$\sigma$ uncertainty values (also displayed above/below lines).  Young ($<$1 Gyr, dark blue) and old ($>$3 Gyr, dark red) \cite{MeibomMathieu(2005)} circularization period measurements are shown for comparison, with $N_b$ denoting the number of cluster measurements with orbital periods less than 100 days.  \textit{Bottom panels}: Circularization periods for the synthetically-observed binaries, varying the number of binaries $N_b$ drawn from the tidally-circularized population, for different values of $\eta$ (eq.~[\ref{eq:eta}]).  See Appendix~\ref{app:SynthPop} for details.
}
\label{fig:cluster-collection}
\end{figure}

The top panel of Figure~\ref{fig:cluster-collection} presents results from such an exercise. With $N_b = 20$ \citep[similar to the sample size of SBs in a given cluster of][$\sim 10-50$]{MeibomMathieu(2005)}, we find a wide range of results, with $P_{\rm M+M} = 8.4^{+3.5}_{-2.4}$ days,  encompassing all cluster results in Figure~\ref{fig:meibom_dist} to within $2\sigma$ (see also vertical lines). This range contracts as $N_b$ rises. The median values for $P_{\rm env}$ and  $P_{\rm M+M}$ also decrease, indicating a bias for longer circularization periods when the sample size is small.

To see if we can reproduce this trend, we calculate $P_{\rm env}$ and $P_{\rm M+M}$ for simulations of tidally-circularized binaries (see App.~\ref{app:SynthPop} for details).  The bottom three panels of Figure~\ref{fig:cluster-collection} display our results, where $\eta$ determines how strongly the dissipation depends on the binary separation \citep[e.g.][]{Ogilvie(2014),Barker(2020)}:
\begin{equation}
\frac{1}{t_{\rm circ}} = \frac{1}{e} \frac{\der e}{\der t} \propto \Porb^{-\eta}.
\label{eq:eta}
\end{equation}
Because we circularize binaries with ages from 1-10 Gyr, $P_{\rm env}$ gives the circularization period for the youngest binaries, while $P_{\rm M+M}$ fits the average circularization period of the population.  Because the range of binary ages causes the eccentricity distribution to be skewed, these fits are biased to long periods when the sample size is small (low $N_b$).  This trend disappears when a narrow range of binary ages is considered (e.g. ages 1-2 Gyr), implying one may indeed robustly determine the circularization period with only a few dozen binaries \citep{MeibomMathieu(2005)}.  The values of $P_{\rm env}$ and $P_{\rm M+M}$ only differentiate themselves when $N_b$ rises above a few hundred binaries, with large uncertainties when $N_b$ is low.  These trends are all reproduced in the \textit{Kepler}/TESS data, with the $P_{\rm env}$, $P_{\rm M+M}$ measurements preferring lower $\eta$ values (see \S\ref{sec:thry_cc} for implications).

\begin{figure}
\includegraphics[width=\linewidth]{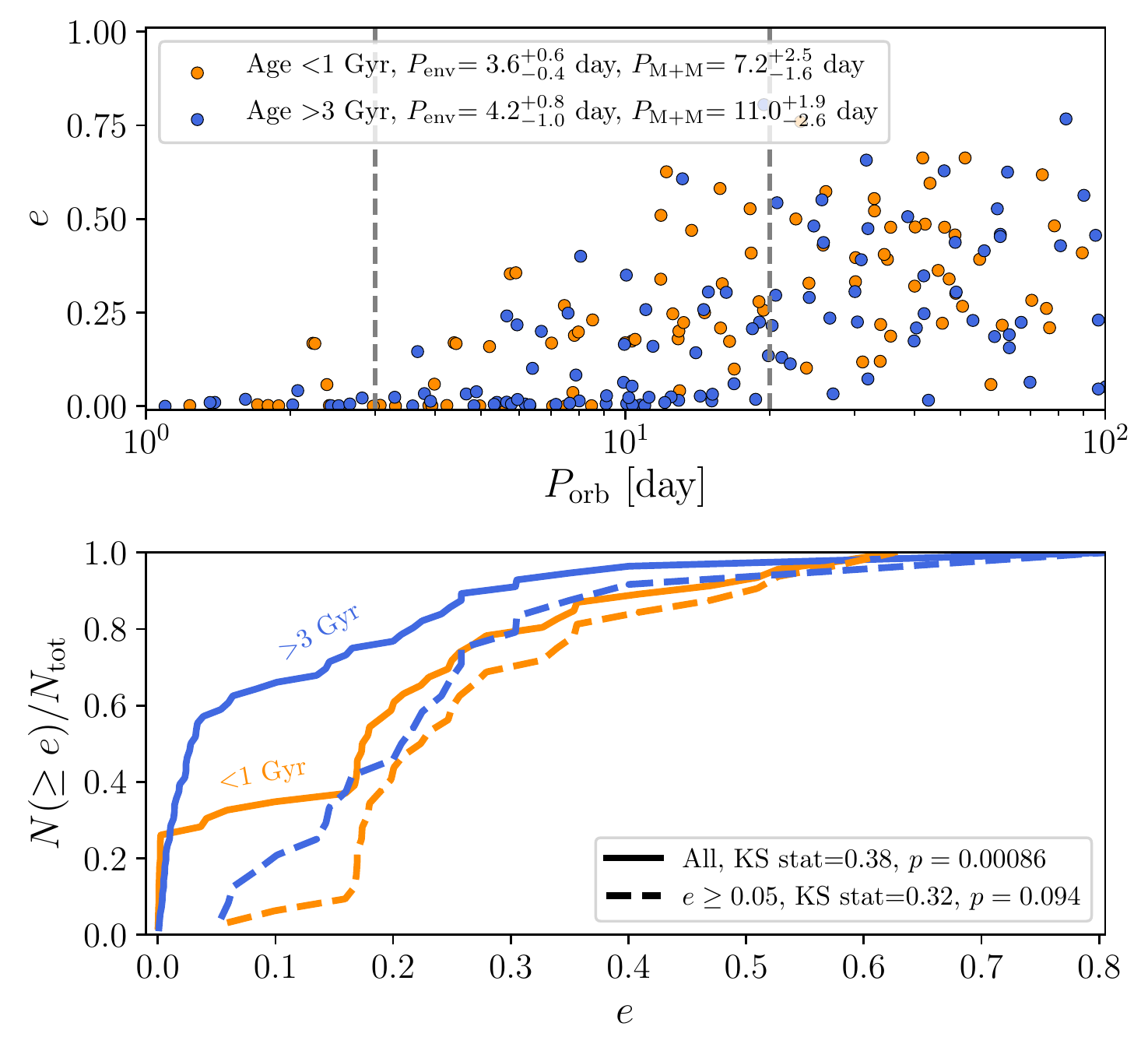}
\caption{
\textit{Top panel:} Young ($<$1 Gyr, orange) and evolved ($>$3 Gyr, blue) binaries from the \cite{MeibomMathieu(2005)} dataset (with updates in M35, \citealt{Leiner(2015)}, and M67 \citealt{Geller(2021)}).  Vertical gray dashed lines delineate orbital periods 3-20 days.  \textit{Bottom panel}: Cumulative eccentricity distribution for the young (orange) and evolved (blue) binaries between 3-20 days, for no eccentricity cutoff (solid), and an  eccentricity cutoff of $e > 0.05$ (dashed).  The legend gives results from a two-sample Kolmogorov–Smirnov (KS) test between the young and old binaries, showing the `eccentricity envelope' ($e>0.05$) is consistent with no evolution with age ($p>0.05$).
}
\label{fig:MM-collection}
\end{figure}

If the \textit{Kepler}/TESS data are detecting multiple circularization periods for binaries with different ages, the younger binaries in the \cite{MeibomMathieu(2005)} data-set should give shorter values for $P_{\rm env}$ and $P_{\rm M+M}$.  Indeed, the top panel of Figure~\ref{fig:cluster-collection} shows young binaries (dark red) analyzed by \cite{MeibomMathieu(2005)} have shorter $P_{\rm M+M}$ values than evolved binaries (dark blue).  However, we now see one can be biased to long $P_{\rm M+M}$ values when $N_b$ is small.

To test the evolution of $P_{\rm env}$ and $P_{\rm M+M}$ with age, we examine the young ($<$1 Gyr) and evolved ($>$3 Gyr) binaries from the \cite{MeibomMathieu(2005)} dataset collectively  (not seperating binaries by cluster, Fig. ~\ref{fig:MM-collection}), adding into consideration new binaries from  \citet{Leiner(2015)} for M35, and from \citet{Geller(2021)} for M67 (see top panel of Fig. \ref{fig:MM-collection}).  The data looks qualitatively very similar to our EB sample, with a clear upper envelope in eccentricity that rises with orbital period, and an over-density of nearly-circular binaries out to beyond $\sim 10$ days.  In the bottom panel of Figure \ref{fig:MM-collection}, we compare the eccentricity distributions for young and old binaries with orbital periods from 3 to 20 days, i.e., the range over which the impact of tidal dissipation is most prominent. A two-sample  Kolmogorov-Smirnov test returns a probability of $p=8.6 \times 10^{-4}$, or it is unlikely that the two distributions are drawn from the same underlying one. Moreover, the older group contains more systems that are circular.

Taken at face value, this would support the \citet{MeibomMathieu(2005)} claim that tidal circularization operates effectively during the main-sequence. However, 
the two populations share the same upper envelope, with  $P_{\rm env} = 3.6^{+0.6}_{-0.4} \ {\rm day}$ for the young sample, and $P_{\rm env} = 4.2^{+0.8}_{-1.0} \ {\rm day}$ for the old ones. The two deviate from each other mostly
in that the older one has a prominent cold core, with $P_{M+M} = 7.2^{+2.5}_{-1.6} \ {\rm day}$ for the young and $P_{\rm M+M} = 11.0^{+1.9}_{-2.6} \ {\rm day}$ for the old.  If we instead only compare binaries with $e \geq 0.05$, i.e., ignoring the cold cores, the KS test returns a $p$-value of $p=0.094$, not significant enough to reject the null hypothesis of drawing from the same distribution ($p < 0.05$).  This casts some doubts on the interpretation of on-going tidal circularization, since $P_{\rm env}$ does not increase with age.

In conclusion, we find that it is hard to robustly determine the circularization period using a small sample of binaries (few tens), due to the presence of the cold core. This explains much of the tension between our work and that of \citet{MeibomMathieu(2005)}. However, the old and young populations in their study do appear to be statistically different, and a proper understanding of this data is required to draw a robust conclusion.

\subsection{The Cold Core and the EBs}

The presence of the cold core group is puzzling.  We posit three possibilities on its formation, giving three separate interpretations for the shortest ($P_{\rm env}$) and average ($P_{\rm M+M}$) circularization period:
\begin{enumerate}
    \item Disk migration and damping formed the core primordially \citep[e.g.][]{Kratter(2010),MoeKratter2019,TokovininMoe(2020)}, implying $P_{\rm env}$ is the only ``true'' measure of tidal circularization.
    \item Inertial waves preferentially synchronized then circularized stars with initially short rotation periods, so $P_{\rm env}$ and $P_{\rm M+M}$ probe two separate tidal dissipation mechanisms.
    \item Different $P_{\rm env}$ and $P_{\rm M+M}$ values probe the circularization of binaries with different ages.
\end{enumerate}
Figure~\ref{fig:MM-collection} tentatively disfavors the third hypothesis, since $P_{\rm env}$ does not seem to increase with binary age.  Reliable age and rotation period data can further illuminate the formation of the cold core.  In \S\ref{sec:thry_cc}, we elaborate on the implications of these interpretations.  

The short-period eccentric binaries probed by the shortest circularization period may also be an abnormality, recently excited by three-body interactions \citep[e.g.][]{MazehShaham(1978),FabryckyTremaine(2007),NaozFabrycky(2014),Hammers(2019),MoeKratter(2018)} or  heart-beat pulsations \citep[e.g.][]{Fuller(2017),ZanazziWu(2020)}.  This would also explain why the young binaries in the \citet{MeibomMathieu(2005)} sample have a less pronounced cold core.  Here we argue this possibility is unlikely, considering the high precision EB data. Plotted in the logarithmic-eccentricity space (right panels of Fig.~\ref{fig:data-ecc-period}), the EB data clearly show the `waterfall' feature, seen also in synthetic observations of eclipsing binaries (right panels of Fig.~\ref{fig:synth-ecc-period}). The eccentricities fall sharply over a narrow period range, since the tidal circularization timescale drops steeply with orbital period. This strong feature of tidal dissipation is difficult to replicate through other processes.
 
Compared to spectroscopic binaries that have more crudely measured eccentricities, EB data are exquisite, and have the power to constrain the process of tidal circularization. In the following, we briefly review a few previous works that have relationship to our study here.

\subsection{Other Previous Works}

A number of recent studies have called 
into question the principal results of \citet{MeibomMathieu(2005)}. Studying field spectroscopic binaries (typically a few Gyrs old) that are of order a few hundred in sample size,  \citet{Milliman(2014),Triaud(2017)} found values of $6-8$ days for the circularization period, in-between that of ours and that of \citet{MeibomMathieu(2005)}.
Similarly, \citet{Price-Whelan(2020),Kounkel} also noticed that many binaries with very short orbital periods still have substantial eccentricities.\footnote{Some of these may be fictitious, see \cite{Price-Whelan(2020)} for a discussion.}

Other studies have explored the eccentricity distributions of EBs, as discovered by \textit{Kepler} and TESS, the very sample we adopt here \citep{VanEylen(2016),JustesenAlbrecht(2021), Kjurkchieva(2017)}. All noticed the significant presence of eccentric binaries within $10$ days, in contradiction to a longer circularization period.
Some of these studies have also investigated the dependency on stellar mass or effective temperature, reaching sometimes diverging results. For instance, while \citet{Torres(2010),VanEylen(2016),JustesenAlbrecht(2021)} found that stars below the Kraft break \citep{Kraft(1967)} appear to be circularized out to longer periods, 
 other works fail to find this trend \citep{Kjurkchieva(2017),Windemuth(2019)}. In this work, we focus solely on solar-type binaries, both because there are more debates around these stars, and because the EBs from {\it Kepler} and  TESS are mostly concentrated in this narrow mass range. Compared to previous studies, we show here that the exquisite precision of eccentricity measurements in EBs can be used to its full potential, and we identify that the observed systems have a broad range of eccentricities, potentially departing from a purely circularizing population (the cold core).
 
\subsection{Implications for Theories of Tidal Dissipation}
\label{sec:Discuss_Thry}

Our work reaches differing conclusions than \cite{MeibomMathieu(2005)}, whose constraints have guided theoretical studies of tidal dissipation in stars for close to two decades.  We first discuss the implications of the eccentricity envelope which circularizes interior to $\sim 3$ days, then the two circularization periods of $\sim 3$ and $\sim 6$ days, for tidal theories.

\subsubsection{The Eccentricity Envelope}
\label{sec:thry_ecc}

A primary contender to circularisation of solar-like binaries is the dissipation of the equilibrium tide by convective turbulence \citep[e.g.][]{Zahn(1989)}. \citet{ZahnBouchet(1989)} have estimated that binaries should be circularized out to $\sim 8$ days after the pre-main-sequence, gradually rising with time outward during the main-sequence. This assumes that the equilibrium tide is efficiently dissipated in the surface convection zones, with the magnitude of turbulent viscosity reduced in the fast-tide regime by a factor of $(P_{\rm orb}/\tau_{\rm cv})^\xi$ and $\xi=1$. Here,
  $\tau_{\rm cv} \gg P_{\rm orb}$ is the characteristic convection turnover time.
 However, multiple works have instead advocated for a much steeper reduction of $\xi = 2$ for fast-tides \citep{GoldreichNicholson(1977),GoodmanOh(1997)}. If so, the equilibrium tide can only circularize  binaries out to $\sim 2$ days during the pre-main-sequence, and little beyond that during the main-sequence \citep{GoodmanOh(1997),GoodmanDickson(1998),Barker(2020),ZanazziWu(2020)}.  Recently, \cite{Terquem(2021a),TerquemMartin(2021)} have argued for un-suppressed dissipation of tidal flows by turbulent convection, arguing that instead of the convective eddies serving as the turbulent viscosity for the tidal flow, the tidal flow works as a viscosity for the turbulent eddies.  They find a circularization period of $\sim$6 days during the pre-main sequence, which increases by $\sim$1-2 days during the main sequence.   The shortest circularization period indirectly supports the most pessimistic estimates on the efficiency of convective damping \citep{GoldreichNicholson(1977),GoodmanDickson(1998)}, in agreement with recent hydrodynamical simulations \citep{OgilvieLesur(2012),VidalBarker(2020a),VidalBarker(2020b),Duguid(2020a),Duguid(2020b),BarkerAstoul(2021)}.

Our results also have bearing on the character of dynamical tides. While dynamical tides without locking have been known to be ineffectual \citep[e.g.][]{Terquem(1998)}, resonance locking can greatly prolong the duration of resonances between tidal forcing and stellar internal modes. This, as calculated by \citet{ZanazziWu(2020)}, can circularize solar-type binaries out to $\sim3-4$ days over the first few million years (pre-main-sequence). However, they found that resonance locking do not operate as efficiently during the main-sequence, when the tidal resonances become too weak. The fact that $P_{\rm env} \sim 3$ days for the \textit{Kepler} and TESS field binaries supports resonance locking operating during the pre-main sequence.

\citet{GoodmanDickson(1998),OgilvieLin(2007),BarkerOgilvie(2010),BarkerOgilvie(2011),Barker(2020)} have pointed out the importance of nonlinear wave-breaking in enhancing the effectiveness of dynamical tides in main-sequence stars. As the radiative cores of these stars are strongly stratified, tidally excited gravity-waves can grow in amplitude as they travel inward. If they overturn, they can deposit all their energy in the stellar core. \citet{GoodmanDickson(1998)} estimated that this can, over the main-sequence lifetime,  circularize solar-type binaries out to $\sim$4-6 days. As this is only a modest increase of the shortest circularization period ($P_{\rm env} \sim$3 days), we could not confirm it using current EB data. We hope that the accumulation of a large sample of binaries, and reasonably precise main-sequence age-dating, will allow us to draw firm conclusions in the future.

\subsubsection{Two Circularization Periods}
\label{sec:thry_cc}

\begin{table}
\begin{center}
\begin{tabular}{|l|c|}
\hline
 Tidal Theory & $\eta$ \\
 \hline 
 Convection Zone Damping$^{1,2,3,4,5}$ & 3.33--6.05 \\
 Non-Resonant Radiative Diffusion$^{1,4}$ & 7 \\
 Non-Resonant Inertial Waves$^{6,4}$ & 6.33 \\
 Non-linear Wave Breaking$^{7,8}$ & 7 \\
\hline
\end{tabular}
\end{center}
\caption{
Tidal theory values for $\eta$ (eq.~[\ref{eq:eta}]).  We require $\eta\lesssim 2.4$ to explain the circularization periods in the \textit{Kepler}/TESS data, not predicted by any tidal theory.  References: 1) \cite{Zahn(1977)}, 2) \cite{GoldreichNicholson(1977)}, 3) \cite{GoodmanOh(1997)}, 4) \cite{Barker(2020)}, 5) \cite{Terquem(2021a)}, 6) \cite{OgilvieLin(2007)}, 7) \cite{GoodmanDickson(1998)}, 8) \cite{BarkerOgilvie(2010)}
}
\label{tab:tides}
\end{table}

An important puzzle arises from our work: the presence of circular binaries out to $\sim 10-20$ days, which we call the `cold core.'  We argue this feature dominates the determination of $P_{\rm M+M} \sim 6 \ {\rm days}$ for the \textit{Kepler}/TESS eclipsing binaries.  This overdensity of circular binaries may be a feature of binary formation, implying $P_{\rm env}$ is the only `true' measure of tidal circularization. It has been argued that solar-type close binaries (inward of 10 AU) are likely the result of disk fragmentation \citep[see, e.g.][]{Kratter(2010),MoeKratter2019,KuruwitaFederrath(2019),TokovininMoe(2020),Kuruwita(2020)}, hence these binaries would be subject to eccentricity damping from their nascent disks. On the other hand, such a scenario fails to explain why many close binaries remain eccentric (to subsequently be tidally circularized). It also can't explain why binaries hosing circumbinary disks are often eccentric \citep[e.g.][]{Czekala(2019)}.

The cold core could also have a primordial tidal origin, but this requires a mechanism which selectively circularizes only a subset of solar-type binaries.  One possibility is circularization via inertial wave dissipation.  The diversity of pre-main sequence rotation periods \citep[e.g.][]{Bouvier(2013)} would allow inertial waves to selectively synchronize then circularize stars born with rapid rotation rates.  However, it is unclear if inertial waves can circularize solar-type stars to $\sim$10 day orbital periods \citep[e.g.][]{OgilvieLin(2007),Barker(2020)}.

If the eccentricity envelope and cold core are the result of age-dependent circularization, the tidal dissipation mechanism cannot depend strongly on the binary separation.  Letting $t_{\rm young}$ ($t_{\rm old}$) be the ages, and $P_{\rm young}$ ($P_{\rm old}$) the circularization period, of the young (old) binaries, equation~\eqref{eq:eta} gives
\be
\frac{t_{\rm young}}{t_{\rm old}} = \left( \frac{P_{\rm young}}{P_{\rm old}} \right)^\eta.
\ee
The \textit{Kepler} and TESS eclipsing binary data requires $\eta \lesssim 2.4$ to explain the two circularization periods ($P_{\rm young} \lesssim P_{\rm env}$, $P_{\rm old} \gtrsim P_{\rm M+M}$),  assuming $t_{\rm young} \approx 1 \ {\rm Gyr}$ and $t_{\rm old} \approx 5 \ {\rm Gyr}$.  Table~\ref{tab:tides} lists $\eta$ values for different tidal theories: all have $\eta$ values significantly larger than $2.4$.  Thus, although no existing tidal theory can give rise to the two circularization periods, a tidal origin cannot be excluded.

\section{Conclusions}
\label{sec:Conc}

In this work, we employ $\sim 500$ eclipsing binaries discovered by the {\it Kepler} and TESS missions to constrain the process of tidal dissipation. With exquisite measurements on the eccentricities (encompassing a dynamic range of $\sim 10^3$), EBs are unique and powerful tools for this goal.

We find at least two circularization periods in the data: a short circularization period of $\sim 3$ days, and an average circularization period of $\sim 6$ days.  The latter circularization period is strongly affected by presence of many nearly-circular binaries out to $\sim$10-20 day orbital periods (the `cold core').  We posit three scenarios to generate these two circularization periods in the data: primordial formation of the `cold core' by disk migration and damping, selective circularization via inertial waves, and age-dependent circularization of field binaries.  These findings are in direct tension with results reported by \citet{MeibomMathieu(2005)}, who studied spectroscopic binaries collected from various open clusters. We point out that the presence of the `cold core' population, together with a much smaller sample size per cluster, may have explained much of the discrepancies between our results and theirs.  However, we reaffirm that their data do show a significant difference between the eccentricity distributions of young ($< 1$ Gyrs) and old ($> 3$ Gyrs) binaries.  More data is needed for a stronger conclusion. Of particular benefit would be many more binaries from the pre-main-sequence phase, such as those collected by  \citet{Melo(2001)} and \citet{Ismailov(2014)}, and accurate age constraints for evolved binaries.

Our results, if confirmed, have the potential to reconcile tidal theories with observations. Assuming the fast-tide reduction supported by recent hydrodynamical simulations, equilibrium tides are not expected to play a significant role in both the main-sequence and the pre-main-sequence.\footnote{We note that dissipation of red giant binaries is well explained by turbulent damping of the equilibrium tide \citet{VerbuntPhinney,PW-goodman}. However, these are not in the `fast-tide' limit.}
First-principle calculations of resonance locking find solar-type binaries circularize out to $\sim 3$ days before they arrive at the main-sequence.
Circularization by resonance locking is consistent with the shortest circularization period found in this work. We cannot, at the moment, exclude a modest rise of the circularization period during the main-sequence, as predicted by theories of wave-breaking. More eclipsing binaries with well defined ages will be needed to answer this.

Lastly, while the tidal process in solar-type binaries may be close to a resolution, there remains much beyond. For instance, stars with radiative envelopes may experience different tidal physics \citep[e.g.][]{Zahn(1975),SavonijePapaloizou(1983),GoldreichNicholson(1989),SuLai(2021)}. EB data from OGLE and other surveys may provide useful constraints \citep[][for LMC and SMC]{ogle-lmc,ogle-stream}. 


\acknowledgments

I thank Yanqin Wu for the significant effort she put into this project, in both the analysis and interpretation of the data.  I thank Robert Mathieu for his constructive criticism which significantly improved the quality of this work, and Simon Albrecht, Katie Breivik, Nathan Hara, Juna Kollmeier,  Maxwell Moe, Norman Murray, Adrian Price-Whelan, Scott Tremaine, Amaury Triaud, Joshua Winn, and Wei Zhu for helpful conversations. 
JZ was supported by the Natural Sciences and Engineering
Research Council of Canada (NSERC) under the funding reference
\# CITA 490888-16.

\appendix

\section{Eccentricity Vectors}
\label{app:ecc}

Here, we briefly review how the eccentricities of eclipsing binaries are measured \citep[see][for a pedagogical review]{Winn(2010)}, and justify our procedure of using one component of the eccentricity vector to constrain tidal evolution. 

Consider an eclipsing binary with the primary and secondary eclipses occurring at time $t_p$ and $t_s$, lasting a duration $T_p$ and $T_s$, respectively. Let the pericentre angle relative to the line-of-sight be $\omega$. We have
\begin{align}
t_s - t_p &\simeq \frac{\Porb}{2} \left(1 + \frac{4}{\pi} e \cos \om \right), \\
\frac{T_s}{T_p} &\simeq \frac{1 + e \sin \om}{1 - e \sin \om},
\end{align}
when $e \ll 1$. These allow both the $e\cos\omega$ and $e\sin\omega$ components to be measured from the lightcurve.

However, while the transit centroids ($t_s$, $t_p$) can be accurately determined (with a similar precision as for the orbital period), the measurements of $T_s$ and $T_p$ are much less precise -- they depend on factors like Limb darkening, observational cadence and stellar noise  \citep[e.g.][]{VanEylen(2016),Windemuth(2019),JustesenAlbrecht(2021)}. As a result, the values of $e\cos\omega$ are typically known much better than those of $e\sin\omega$. This explains the strange clustering seen in Figure \ref{fig:period_pomega}. In order to utilize the full potential of EB data, we have therefore opted to focus only on the  $e\cos\omega$ measurements.

\section{Synthetic Observations of Circularized Binaries}
\label{app:SynthPop}

To understand how the sample size impacts measurements of the circularization period, we produce synthetic observations of tidally-circularized binaries.  Given an initial binary eccentricity $e$ and orbital period $\Porb$, we assume pseudo-synchronous rotation, and evolve the binary orbit as \citep{Hut(1981)}
\begin{align}
    &\frac{1}{e} \frac{\der e}{\der t} = \frac{\mu(1+\mu)}{t_c} F_e(e),
    \label{eq:dedt}\\
    &\frac{1}{\Porb} \frac{\der \Porb}{\der t} = \frac{3 \mu(1+\mu)}{2 t_c} F_a(e),
    \label{eq:dPdt}
\end{align}
where $\mu\le1$ is the mass ratio,
\begin{align}
    F_e(e) &= \frac{\Omega_e(e) N(e)}{\Omega(e)} - \frac{18}{11} N_e(e),
    \label{eq:Fe}\\
    F_a(e) &= \frac{4}{11} \left[ \frac{N^2(e)}{\Omega(e)} - N_a(e) \right],
\end{align}
and the functions $N(e)$, $\Omega(e)$, $\Omega_e(e)$, $N_a(e)$, and $N_e(e)$ are defined in \cite{Leconte(2010)}.  We parameterize the circularization period as 
\begin{equation}
    t_c = 0.3 \ {\rm Gyr} \left( \frac{\Porb}{4 \ {\rm day}} \right)^\eta,
\end{equation}
with $\eta$ a free parameter which governs how strongly $t_c$ depends on the binary separation (see Table~\ref{tab:tides} for physical $\eta$ values).  

\begin{figure}
\includegraphics[width=\linewidth]{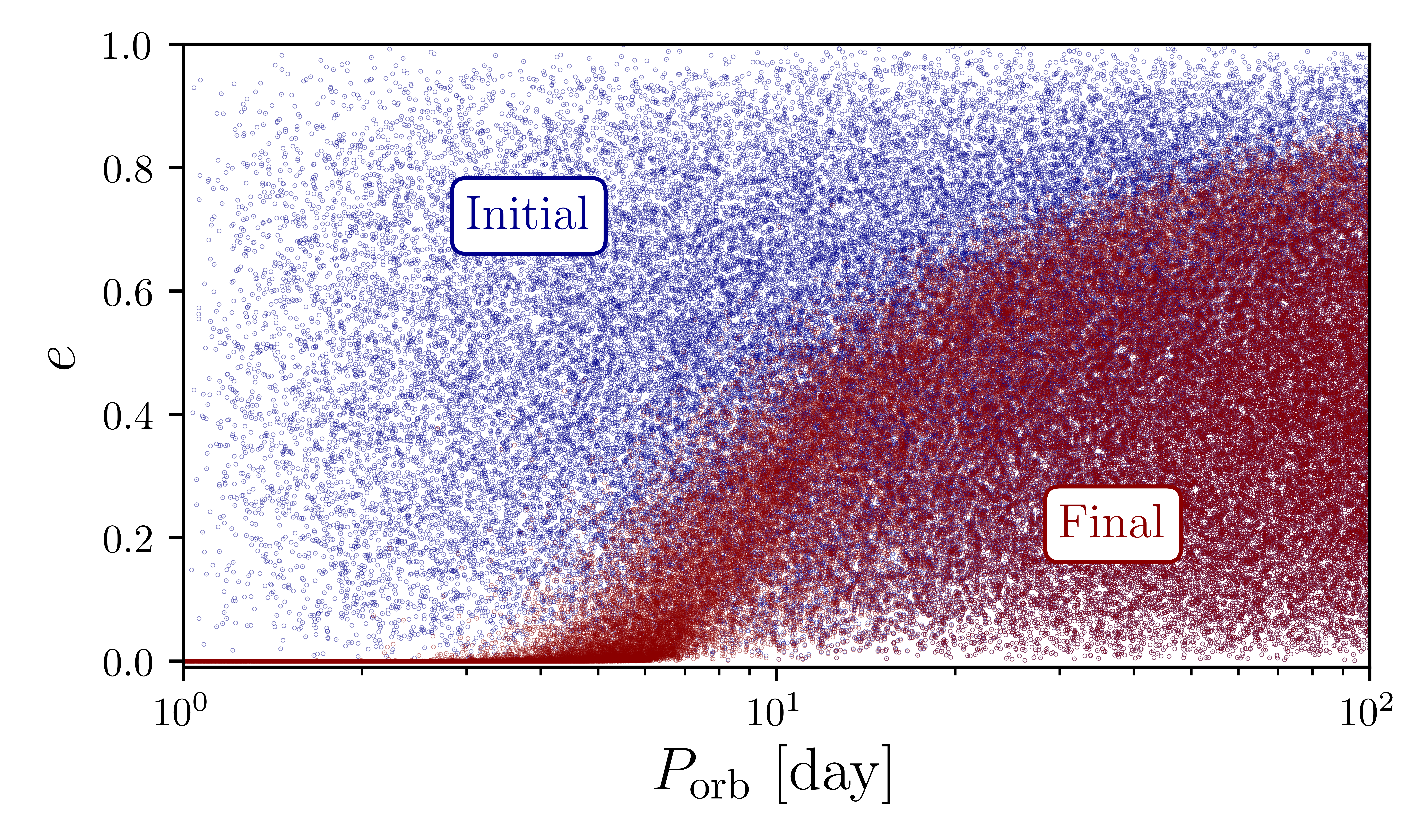}
\caption{
Initial (blue) and final (red) population of binaries after tidal circualrization, for $\eta=4.5$.  See text for details.
}
\label{fig:single-pop}
\end{figure}

Our initial distribution of $e$, $\Porb$, and $\mu$ values are motivated by observations.  We draw eccentricity values from a beta distribution $\mathcal{B}(e|a,b)$, with $a=1.75$ and $b=2.01$, constrained from the APOGEE Gold Sample \citep{Price-Whelan(2020)}.  Mass ratios are drawn from a linear distribution $\mathcal{P}(\mu) \propto \mu$, to mimic the abundance of equal-mass binaries at short periods \citep[e.g.][]{Raghavan(2010),MoeDiStefano(2017),Windemuth(2019)}.  To approximate the binary period log-normal distribution centered at $\Porb \sim 250 \ {\rm yrs}$ \citep{Raghavan(2010)}, we assign initial periods through
\begin{equation}
    \ln \Porb = \lambda \ln P_{\rm min} + (1-\lambda) \ln P_{\rm max},
\end{equation}
with $0 \le \lambda \le 1$ drawn from a linear distribution $\mathcal{P}(\lambda) \propto \lambda$, with $P_{\rm min} = 1 \ {\rm day}$ and $P_{\rm max} = 200 \ {\rm day}$.  We integrate equations~\eqref{eq:dedt}-\eqref{eq:dPdt} for $N_{\rm tot} = 10^5$ binaries from $t=0$ to the binary age $t=t_{\rm age}$, uniformly distributed between 1-10 Gyr.  Figure~\ref{fig:single-pop} displays the initial and final population for one of our simulations.

\begin{figure*}
\includegraphics[width=\textwidth]{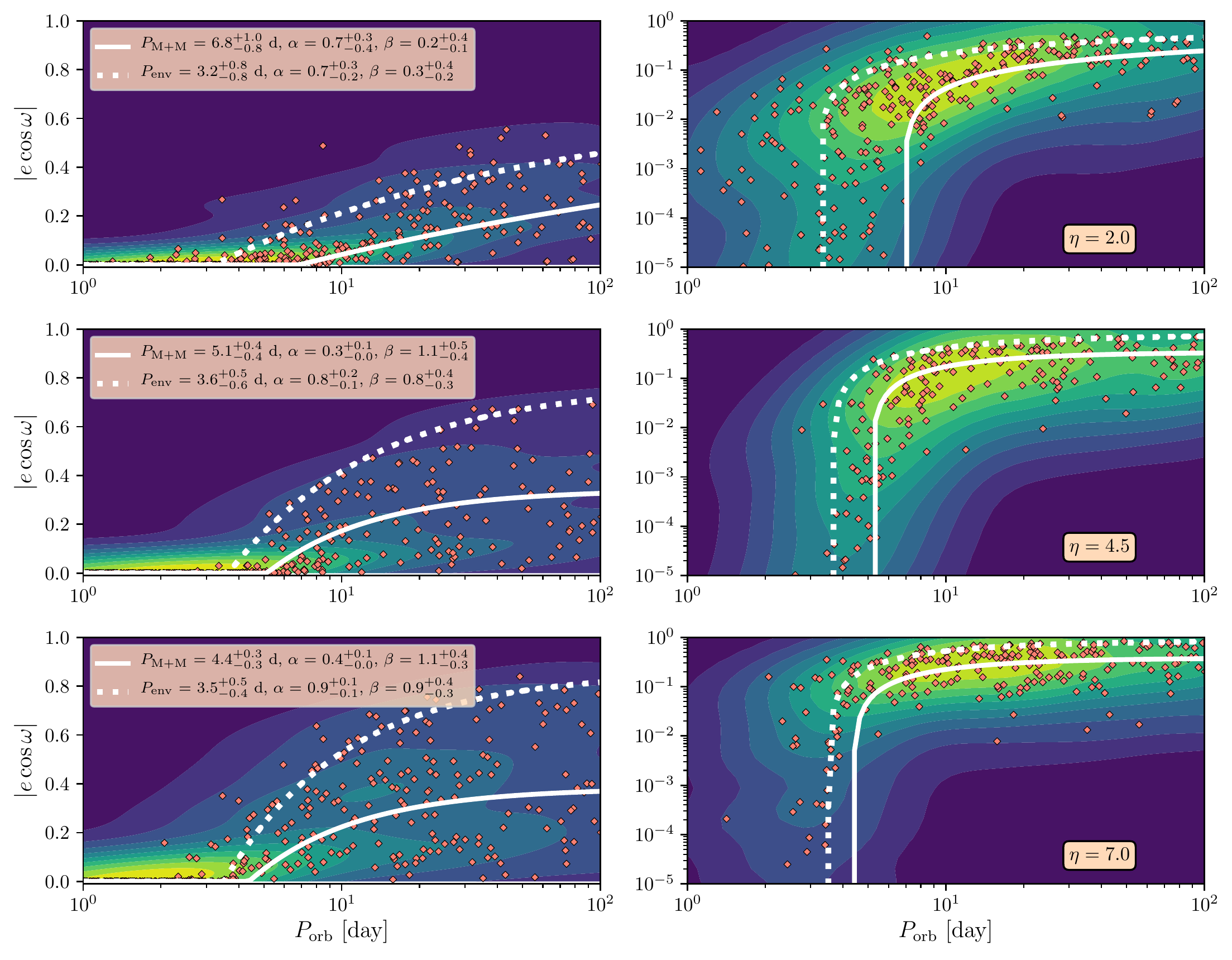}
\caption{
Synthetic Observations of $N_b = 524$ tidally-circularized binaries, for the $\eta$ values indicated.  Salmon diamonds denote projected eccentricity measurements, contours denote smoothed Kernel Density Estimates of the projected eccentricies, while solid (dotted) lines denote $P_{\rm M+M}$ ($P_{\rm env}$) fits to 1000 observations of the circularized population.  The $P_{\rm M+M}$, $P_{\rm env}$ parameter fits are displayed in the legends.
}
\label{fig:synth-ecc-period}
\end{figure*}

To generate a synthetic observation, we assign each binary a longitude of pericenter $\om$, distributed uniformly between $0$ to $2\pi$.  Projected eccentricities $|e \cos \omega|$ are measured by drawing $N_b$ binaries from the theoretical population, weighted by the probability primary and secondary eclipses are observed \citep[e.g.][]{Winn(2010)}:
\begin{equation}
    p_{\rm ecl} \propto \frac{1}{a} \left( \frac{1 - |e \sin \om|}{1 - e^2} \right).
\end{equation}
We calculate $P_{\rm M+M}$ and $P_{\rm env}$ for the $N_b$ eclipsing binaries, and repeat this process $10^3$ times.

\begin{figure}
\includegraphics[width=\linewidth]{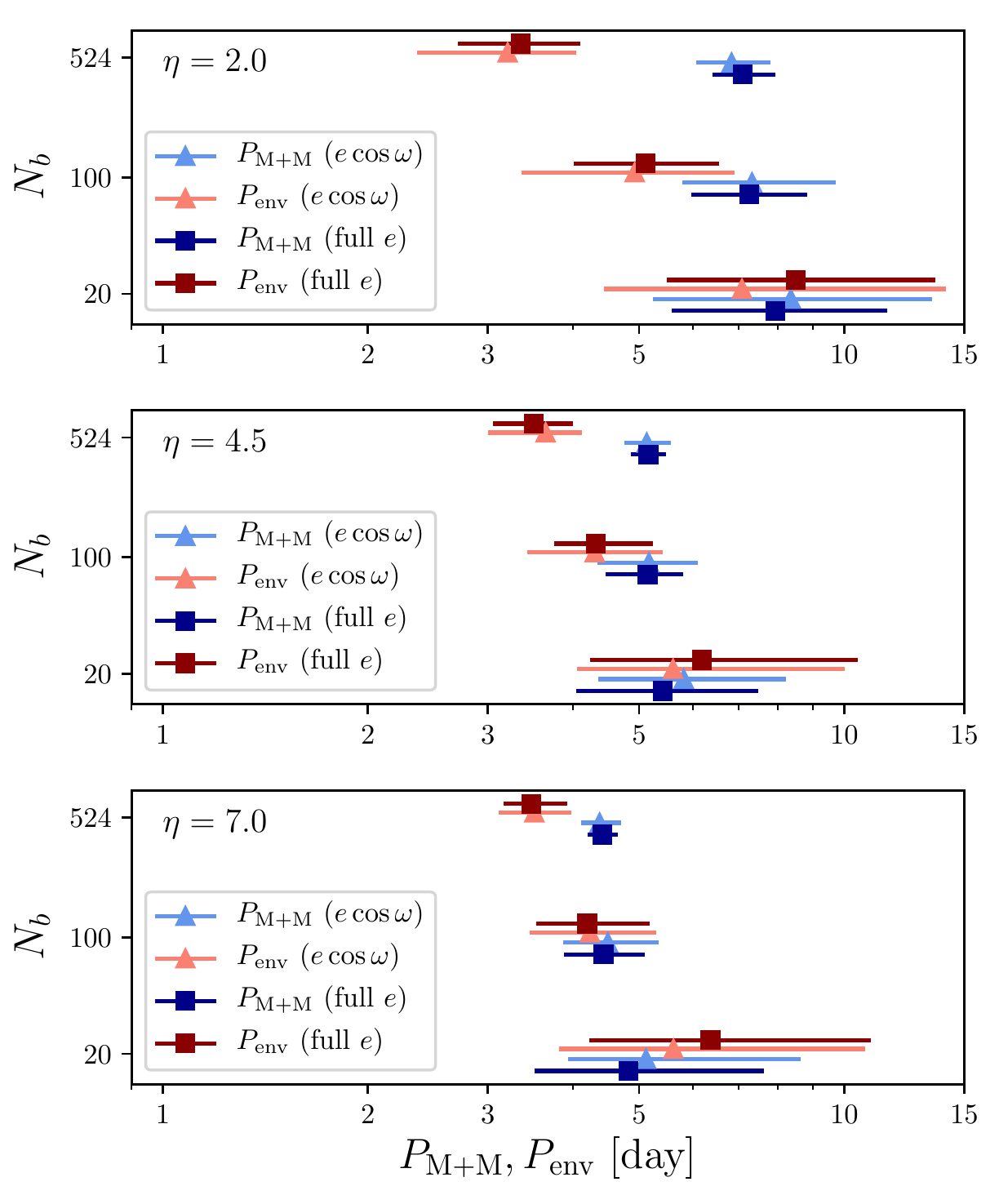}
\caption{
Comparing $P_{\rm M+M}$, $P_{\rm env}$ constraints using the projected (triangles) and full (squares) eccentricity measurements, drawing $N_b$ binaries from the tidally-circularized population, for the $\eta$ values indicated.  Lines denote 1$\sigma$ errors, slightly displaced from their $y$-axis $N_b$ value for clarity.  There is no notable difference between $P_{\rm M+M}$, $P_{\rm env}$ constraints using the projected vs. full eccentricity measurements.
}
\label{fig:proj-check}
\end{figure}

Our simulations find a stronger dependence of eccentricity damping on orbital period leads to a more narrow range of circularization periods.  Figure~\ref{fig:synth-ecc-period} displays synthetic observations of our simulations, alongside constraints on the circularization period.  We see only low values of $\eta$ can lead to significantly different circularization periods.  We also see tidal dissipation can lead to a wide range of very small eccentricity values at short orbital periods, seen also in the eclipsing binary data (the `waterfall' in Fig.~\ref{fig:data-ecc-period}).

However, a natural question is if the circularization period constraint differs if you only use the projected eccentricities from eclipsing binaries, or the full eccentricities from spectroscopic binaries.  To check if keeping only the projected eccentricities alters the determination of $P_{\rm M+M}$ or $P_{\rm env}$, we compare our eclipsing binary $P_{\rm env}$ and $P_{\rm M+M}$ values to synthetic spectroscopic binary values.   To create an observation of spectroscopic binaries, the orbits of $N_b$ binaries are drawn randomly from our simulation, and then equation~\eqref{eq:fe} is fit to the full eccentricity and orbital period data.  Figure~\ref{fig:proj-check} displays the results of this calculation, where find no significant difference in $P_{\rm env}$ and $P_{\rm M+M}$ values when fit to eclipsing or spectroscopic binaries.

\bibliographystyle{aasjournal}
\bibliography{main}

\begin{thebibliography}{}
\expandafter\ifx\csname natexlab\endcsname\relax\def\natexlab#1{#1}\fi
\providecommand{\url}[1]{\href{#1}{#1}}

\bibitem[{{Barker}(2020)}]{Barker(2020)}
{Barker}, A.~J. 2020, \mnras, arXiv:2008.03262

\bibitem[{{Barker} \& {Astoul}(2021)}]{BarkerAstoul(2021)}
{Barker}, A.~J., \& {Astoul}, A. A.~V. 2021, arXiv e-prints, arXiv:2105.00757

\bibitem[{{Barker} \& {Ogilvie}(2010)}]{BarkerOgilvie(2010)}
{Barker}, A.~J., \& {Ogilvie}, G.~I. 2010, \mnras, 404, 1849

\bibitem[{{Barker} \& {Ogilvie}(2011)}]{BarkerOgilvie(2011)}
---. 2011, \mnras, 417, 745

\bibitem[{{Borucki} {et~al.}(2010){Borucki}, {Koch}, {Basri}, {Batalha},
  {Brown}, {Caldwell}, {Caldwell}, {Christensen-Dalsgaard}, {Cochran},
  {DeVore}, {Dunham}, {Dupree}, {Gautier}, {Geary}, {Gilliland}, {Gould},
  {Howell}, {Jenkins}, {Kondo}, {Latham}, {Marcy}, {Meibom}, {Kjeldsen},
  {Lissauer}, {Monet}, {Morrison}, {Sasselov}, {Tarter}, {Boss}, {Brownlee},
  {Owen}, {Buzasi}, {Charbonneau}, {Doyle}, {Fortney}, {Ford}, {Holman},
  {Seager}, {Steffen}, {Welsh}, {Rowe}, {Anderson}, {Buchhave}, {Ciardi},
  {Walkowicz}, {Sherry}, {Horch}, {Isaacson}, {Everett}, {Fischer}, {Torres},
  {Johnson}, {Endl}, {MacQueen}, {Bryson}, {Dotson}, {Haas}, {Kolodziejczak},
  {Van Cleve}, {Chandrasekaran}, {Twicken}, {Quintana}, {Clarke}, {Allen},
  {Li}, {Wu}, {Tenenbaum}, {Verner}, {Bruhweiler}, {Barnes}, \&
  {Prsa}}]{Borucki+(2010)}
{Borucki}, W.~J., {Koch}, D., {Basri}, G., {et~al.} 2010, Science, 327, 977

\bibitem[{{Bouvier}(2013)}]{Bouvier(2013)}
{Bouvier}, J. 2013, in EAS Publications Series, Vol.~62, EAS Publications
  Series, ed. P.~{Hennebelle} \& C.~{Charbonnel}, 143--168

\bibitem[{{Burkart} {et~al.}(2012){Burkart}, {Quataert}, {Arras}, \&
  {Weinberg}}]{Burkart(2012)}
{Burkart}, J., {Quataert}, E., {Arras}, P., \& {Weinberg}, N.~N. 2012, \mnras,
  421, 983

\bibitem[{{Czekala} {et~al.}(2019){Czekala}, {Chiang}, {Andrews}, {Jensen},
  {Torres}, {Wilner}, {Stassun}, \& {Macintosh}}]{Czekala(2019)}
{Czekala}, I., {Chiang}, E., {Andrews}, S.~M., {et~al.} 2019, \apj, 883, 22

\bibitem[{{Duguid} {et~al.}(2020{\natexlab{a}}){Duguid}, {Barker}, \&
  {Jones}}]{Duguid(2020a)}
{Duguid}, C.~D., {Barker}, A.~J., \& {Jones}, C.~A. 2020{\natexlab{a}}, \mnras,
  497, 3400

\bibitem[{{Duguid} {et~al.}(2020{\natexlab{b}}){Duguid}, {Barker}, \&
  {Jones}}]{Duguid(2020b)}
---. 2020{\natexlab{b}}, \mnras, 491, 923

\bibitem[{{Fabrycky} \& {Tremaine}(2007)}]{FabryckyTremaine(2007)}
{Fabrycky}, D., \& {Tremaine}, S. 2007, \apj, 669, 1298

\bibitem[{{Fuller}(2017)}]{Fuller(2017)}
{Fuller}, J. 2017, \mnras, 472, 1538

\bibitem[{{Fuller} \& {Lai}(2012)}]{FullerLai(2012)}
{Fuller}, J., \& {Lai}, D. 2012, \mnras, 420, 3126

\bibitem[{{Geller} {et~al.}(2013){Geller}, {Hurley}, \&
  {Mathieu}}]{Geller+(2013)}
{Geller}, A.~M., {Hurley}, J.~R., \& {Mathieu}, R.~D. 2013, \aj, 145, 8

\bibitem[{{Geller} \& {Mathieu}(2012)}]{Geller+(2012)}
{Geller}, A.~M., \& {Mathieu}, R.~D. 2012, \aj, 144, 54

\bibitem[{{Geller} {et~al.}(2021){Geller}, {Mathieu}, {Latham}, {Pollack},
  {Torres}, \& {Leiner}}]{Geller(2021)}
{Geller}, A.~M., {Mathieu}, R.~D., {Latham}, D.~W., {et~al.} 2021, \aj, 161,
  190

\bibitem[{{Goldreich} \& {Nicholson}(1977)}]{GoldreichNicholson(1977)}
{Goldreich}, P., \& {Nicholson}, P.~D. 1977, \icarus, 30, 301

\bibitem[{{Goldreich} \& {Nicholson}(1989)}]{GoldreichNicholson(1989)}
---. 1989, \apj, 342, 1079

\bibitem[{{Goodman} \& {Dickson}(1998)}]{GoodmanDickson(1998)}
{Goodman}, J., \& {Dickson}, E.~S. 1998, \apj, 507, 938

\bibitem[{{Goodman} \& {Lackner}(2009)}]{GoodmanLackner}
{Goodman}, J., \& {Lackner}, C. 2009, \apj, 696, 2054

\bibitem[{{Goodman} \& {Oh}(1997)}]{GoodmanOh(1997)}
{Goodman}, J., \& {Oh}, S.~P. 1997, \apj, 486, 403

\bibitem[{{Hamers}(2019)}]{Hammers(2019)}
{Hamers}, A.~S. 2019, \mnras, 482, 2262

\bibitem[{{Hut}(1981)}]{Hut(1981)}
{Hut}, P. 1981, \aap, 99, 126

\bibitem[{{Ismailov} {et~al.}(2014){Ismailov}, {Abdi}, \&
  {Mamedxanova}}]{Ismailov(2014)}
{Ismailov}, N.~Z., {Abdi}, H.~A., \& {Mamedxanova}, G.~B. 2014,
  Astronomicheskij Tsirkulyar, 1610, 1

\bibitem[{{Justesen} \& {Albrecht}(2021)}]{JustesenAlbrecht(2021)}
{Justesen}, A.~B., \& {Albrecht}, S. 2021, arXiv e-prints, arXiv:2103.09216

\bibitem[{{Kirk} {et~al.}(2016){Kirk}, {Conroy}, {Pr{\v{s}}a}, {Abdul-Masih},
  {Kochoska}, {Matijevi{\v{c}}}, {Hambleton}, {Barclay}, {Bloemen}, {Boyajian},
  {Doyle}, {Fulton}, {Hoekstra}, {Jek}, {Kane}, {Kostov}, {Latham}, {Mazeh},
  {Orosz}, {Pepper}, {Quarles}, {Ragozzine}, {Shporer}, {Southworth},
  {Stassun}, {Thompson}, {Welsh}, {Agol}, {Derekas}, {Devor}, {Fischer},
  {Green}, {Gropp}, {Jacobs}, {Johnston}, {LaCourse}, {Saetre}, {Schwengeler},
  {Toczyski}, {Werner}, {Garrett}, {Gore}, {Martinez}, {Spitzer}, {Stevick},
  {Thomadis}, {Vrijmoet}, {Yenawine}, {Batalha}, \& {Borucki}}]{kirk}
{Kirk}, B., {Conroy}, K., {Pr{\v{s}}a}, A., {et~al.} 2016, \aj, 151, 68

\bibitem[{{Kjurkchieva} {et~al.}(2017){Kjurkchieva}, {Vasileva}, \&
  {Atanasova}}]{Kjurkchieva(2017)}
{Kjurkchieva}, D., {Vasileva}, D., \& {Atanasova}, T. 2017, \aj, 154, 105

\bibitem[{{Kounkel} {et~al.}(2021){Kounkel}, {Covey}, {Stassun},
  {Price-Whelan}, {Holtzman}, {Chojnowski}, {Longa-Pe{\~n}a},
  {Rom{\'a}n-Z{\'u}{\~n}iga}, {Hernandez}, {Serna}, {Badenes}, {De Lee},
  {Majewski}, {Stringfellow}, {Kratter}, {Moe}, {Frinchaboy}, {Beaton},
  {Fern{\'a}ndez-Trincado}, {Mahadevan}, {Minniti}, {Beers}, {Schneider},
  {Barb{\'a}}, {Brownstein}, {An{\'\i}bal Garc{\'\i}a-Hern{\'a}ndez}, {Pan}, \&
  {Bizyaev}}]{Kounkel}
{Kounkel}, M., {Covey}, K.~R., {Stassun}, K.~G., {et~al.} 2021, arXiv e-prints,
  arXiv:2107.10860

\bibitem[{{Kraft}(1967)}]{Kraft(1967)}
{Kraft}, R.~P. 1967, \apj, 150, 551

\bibitem[{{Kratter} {et~al.}(2010){Kratter}, {Matzner}, {Krumholz}, \&
  {Klein}}]{Kratter(2010)}
{Kratter}, K.~M., {Matzner}, C.~D., {Krumholz}, M.~R., \& {Klein}, R.~I. 2010,
  \apj, 708, 1585

\bibitem[{{Kuruwita} \& {Federrath}(2019)}]{KuruwitaFederrath(2019)}
{Kuruwita}, R.~L., \& {Federrath}, C. 2019, \mnras, 486, 3647

\bibitem[{{Kuruwita} {et~al.}(2020){Kuruwita}, {Federrath}, \&
  {Haugb{\o}lle}}]{Kuruwita(2020)}
{Kuruwita}, R.~L., {Federrath}, C., \& {Haugb{\o}lle}, T. 2020, \aap, 641, A59

\bibitem[{{Latham} {et~al.}(1992){Latham}, {Mathieu}, {Milone}, \&
  {Davis}}]{Latham+(1992)}
{Latham}, D.~W., {Mathieu}, R.~D., {Milone}, A.~A.~E., \& {Davis}, R.~J. 1992,
  in Astronomical Society of the Pacific Conference Series, Vol.~32, IAU
  Colloq. 135: Complementary Approaches to Double and Multiple Star Research,
  ed. H.~A. {McAlister} \& W.~I. {Hartkopf}, 152

\bibitem[{{Leconte} {et~al.}(2010){Leconte}, {Chabrier}, {Baraffe}, \&
  {Levrard}}]{Leconte(2010)}
{Leconte}, J., {Chabrier}, G., {Baraffe}, I., \& {Levrard}, B. 2010, \aap, 516,
  A64

\bibitem[{{Leiner} {et~al.}(2015){Leiner}, {Mathieu}, {Gosnell}, \&
  {Geller}}]{Leiner(2015)}
{Leiner}, E.~M., {Mathieu}, R.~D., {Gosnell}, N.~M., \& {Geller}, A.~M. 2015,
  \aj, 150, 10

\bibitem[{{Lin} \& {Ogilvie}(2018)}]{LinOgilvie(2018)}
{Lin}, Y., \& {Ogilvie}, G.~I. 2018, \mnras, 474, 1644

\bibitem[{{Lin} \& {Ogilvie}(2021)}]{LinOgilvie(2021)}
---. 2021, \apjl, 918, L21

\bibitem[{{Ma} \& {Fuller}(2021)}]{MaFuller(2021)}
{Ma}, L., \& {Fuller}, J. 2021, arXiv e-prints, arXiv:2105.09335

\bibitem[{{Mathieu} {et~al.}(2004){Mathieu}, {Meibom}, \&
  {Dolan}}]{Mathieu+(2004)}
{Mathieu}, R.~D., {Meibom}, S., \& {Dolan}, C.~J. 2004, \apjl, 602, L121

\bibitem[{{Mazeh}(2008)}]{Mazeh(2008)}
{Mazeh}, T. 2008, in EAS Publications Series, Vol.~29, EAS Publications Series,
  ed. M.~J. {Goupil} \& J.~P. {Zahn}, 1--65

\bibitem[{{Mazeh} \& {Shaham}(1978)}]{MazehShaham(1978)}
{Mazeh}, T., \& {Shaham}, J. 1978, \astap, 77, 145

\bibitem[{{Meibom} \& {Mathieu}(2005)}]{MeibomMathieu(2005)}
{Meibom}, S., \& {Mathieu}, R.~D. 2005, \apj, 620, 970

\bibitem[{{Meibom} {et~al.}(2006){Meibom}, {Mathieu}, \&
  {Stassun}}]{Meibom+(2006)}
{Meibom}, S., {Mathieu}, R.~D., \& {Stassun}, K.~G. 2006, \apj, 653, 621

\bibitem[{{Melo} {et~al.}(2001){Melo}, {Covino}, {Alcal{\'a}}, \&
  {Torres}}]{Melo(2001)}
{Melo}, C.~H.~F., {Covino}, E., {Alcal{\'a}}, J.~M., \& {Torres}, G. 2001,
  \aap, 378, 898

\bibitem[{{Milliman} {et~al.}(2014){Milliman}, {Mathieu}, {Geller}, {Gosnell},
  {Meibom}, \& {Platais}}]{Milliman(2014)}
{Milliman}, K.~E., {Mathieu}, R.~D., {Geller}, A.~M., {et~al.} 2014, \aj, 148,
  38

\bibitem[{{Moe} \& {Di Stefano}(2017)}]{MoeDiStefano(2017)}
{Moe}, M., \& {Di Stefano}, R. 2017, \apjs, 230, 15

\bibitem[{{Moe} \& {Kratter}(2018)}]{MoeKratter(2018)}
{Moe}, M., \& {Kratter}, K.~M. 2018, \apj, 854, 44

\bibitem[{{Moe} {et~al.}(2019){Moe}, {Kratter}, \& {Badenes}}]{MoeKratter2019}
{Moe}, M., {Kratter}, K.~M., \& {Badenes}, C. 2019, \apj, 875, 61

\bibitem[{{Naoz} \& {Fabrycky}(2014)}]{NaozFabrycky(2014)}
{Naoz}, S., \& {Fabrycky}, D.~C. 2014, \apj, 793, 137

\bibitem[{{Nine} {et~al.}(2020){Nine}, {Milliman}, {Mathieu}, {Geller},
  {Leiner}, {Platais}, \& {Tofflemire}}]{Nine+(2020)}
{Nine}, A.~C., {Milliman}, K.~E., {Mathieu}, R.~D., {et~al.} 2020, \aj, 160,
  169

\bibitem[{{North} \& {Zahn}(2003)}]{NorthZahn(2003)}
{North}, P., \& {Zahn}, J.~P. 2003, \aap, 405, 677

\bibitem[{{Ogilvie}(2014)}]{Ogilvie(2014)}
{Ogilvie}, G.~I. 2014, \araa, 52, 171

\bibitem[{{Ogilvie} \& {Lesur}(2012)}]{OgilvieLesur(2012)}
{Ogilvie}, G.~I., \& {Lesur}, G. 2012, \mnras, 422, 1975

\bibitem[{{Ogilvie} \& {Lin}(2007)}]{OgilvieLin(2007)}
{Ogilvie}, G.~I., \& {Lin}, D.~N.~C. 2007, \apj, 661, 1180

\bibitem[{{Pawlak} {et~al.}(2016){Pawlak}, {Soszy{\'n}ski}, {Udalski},
  {Szyma{\'n}ski}, {Wyrzykowski}, {Ulaczyk}, {Poleski}, {Pietrukowicz},
  {Koz{\l}owski}, {Skowron}, {Skowron}, {Mr{\'o}z}, \&
  {Hamanowicz}}]{ogle-stream}
{Pawlak}, M., {Soszy{\'n}ski}, I., {Udalski}, A., {et~al.} 2016, \actaa, 66,
  421

\bibitem[{{Penev} {et~al.}(2007){Penev}, {Sasselov}, {Robinson}, \&
  {Demarque}}]{Penev(2007)}
{Penev}, K., {Sasselov}, D., {Robinson}, F., \& {Demarque}, P. 2007, \apj, 655,
  1166

\bibitem[{{Price-Whelan} \& {Goodman}(2018)}]{PW-goodman}
{Price-Whelan}, A.~M., \& {Goodman}, J. 2018, \apj, 867, 5

\bibitem[{{Price-Whelan} {et~al.}(2017){Price-Whelan}, {Hogg},
  {Foreman-Mackey}, \& {Rix}}]{Price-Whelan(2017)}
{Price-Whelan}, A.~M., {Hogg}, D.~W., {Foreman-Mackey}, D., \& {Rix}, H.-W.
  2017, \apj, 837, 20

\bibitem[{{Price-Whelan} {et~al.}(2020){Price-Whelan}, {Hogg}, {Rix}, {Beaton},
  {Lewis}, {Nidever}, {Almeida}, {Badenes}, {Barba}, {Beers}, {Carlberg}, {De
  Lee}, {Fern{\'a}ndez-Trincado}, {Frinchaboy}, {Garc{\'\i}a-Hern{\'a}ndez},
  {Green}, {Hasselquist}, {Longa-Pe{\~n}a}, {Majewski}, {Nitschelm}, {Sobeck},
  {Stassun}, {Stringfellow}, \& {Troup}}]{Price-Whelan(2020)}
{Price-Whelan}, A.~M., {Hogg}, D.~W., {Rix}, H.-W., {et~al.} 2020, \apj, 895, 2

\bibitem[{{Pr{\v{s}}a} {et~al.}(2011){Pr{\v{s}}a}, {Batalha}, {Slawson},
  {Doyle}, {Welsh}, {Orosz}, {Seager}, {Rucker}, {Mjaseth}, {Engle}, {Conroy},
  {Jenkins}, {Caldwell}, {Koch}, \& {Borucki}}]{Prsa(2011)}
{Pr{\v{s}}a}, A., {Batalha}, N., {Slawson}, R.~W., {et~al.} 2011, \aj, 141, 83

\bibitem[{{Raghavan} {et~al.}(2010){Raghavan}, {McAlister}, {Henry}, {Latham},
  {Marcy}, {Mason}, {Gies}, {White}, \& {ten Brummelaar}}]{Raghavan(2010)}
{Raghavan}, D., {McAlister}, H.~A., {Henry}, T.~J., {et~al.} 2010, \apjs, 190,
  1

\bibitem[{{Ricker} {et~al.}(2015){Ricker}, {Winn}, {Vanderspek}, {Latham},
  {Bakos}, {Bean}, {Berta-Thompson}, {Brown}, {Buchhave}, {Butler}, {Butler},
  {Chaplin}, {Charbonneau}, {Christensen-Dalsgaard}, {Clampin}, {Deming},
  {Doty}, {De Lee}, {Dressing}, {Dunham}, {Endl}, {Fressin}, {Ge}, {Henning},
  {Holman}, {Howard}, {Ida}, {Jenkins}, {Jernigan}, {Johnson}, {Kaltenegger},
  {Kawai}, {Kjeldsen}, {Laughlin}, {Levine}, {Lin}, {Lissauer}, {MacQueen},
  {Marcy}, {McCullough}, {Morton}, {Narita}, {Paegert}, {Palle}, {Pepe},
  {Pepper}, {Quirrenbach}, {Rinehart}, {Sasselov}, {Sato}, {Seager},
  {Sozzetti}, {Stassun}, {Sullivan}, {Szentgyorgyi}, {Torres}, {Udry}, \&
  {Villasenor}}]{Ricker+(2015)}
{Ricker}, G.~R., {Winn}, J.~N., {Vanderspek}, R., {et~al.} 2015, Journal of
  Astronomical Telescopes, Instruments, and Systems, 1, 014003

\bibitem[{{Savonije} \& {Papaloizou}(1983)}]{SavonijePapaloizou(1983)}
{Savonije}, G.~J., \& {Papaloizou}, J.~C.~B. 1983, \mnras, 203, 581

\bibitem[{{Savonije} \& {Papaloizou}(1984)}]{SavonijePapaloizou(1984)}
---. 1984, \mnras, 207, 685

\bibitem[{{Savonije} \& {Witte}(2002)}]{SavonijeWitte(2002)}
{Savonije}, G.~J., \& {Witte}, M.~G. 2002, \aap, 386, 211

\bibitem[{{Slawson} {et~al.}(2011){Slawson}, {Pr{\v{s}}a}, {Welsh}, {Orosz},
  {Rucker}, {Batalha}, {Doyle}, {Engle}, {Conroy}, {Coughlin}, {Gregg},
  {Fetherolf}, {Short}, {Windmiller}, {Fabrycky}, {Howell}, {Jenkins}, {Uddin},
  {Mullally}, {Seader}, {Thompson}, {Sand erfer}, {Borucki}, \&
  {Koch}}]{Slawson(2011)}
{Slawson}, R.~W., {Pr{\v{s}}a}, A., {Welsh}, W.~F., {et~al.} 2011, \aj, 142,
  160

\bibitem[{{Su} \& {Lai}(2021)}]{SuLai(2021)}
{Su}, Y., \& {Lai}, D. 2021, arXiv e-prints, arXiv:2110.12030

\bibitem[{{Terquem}(2021)}]{Terquem(2021a)}
{Terquem}, C. 2021, \mnras, 503, 5789

\bibitem[{{Terquem} \& {Martin}(2021)}]{TerquemMartin(2021)}
{Terquem}, C., \& {Martin}, S. 2021, \mnras, 507, 4165

\bibitem[{{Terquem} {et~al.}(1998){Terquem}, {Papaloizou}, {Nelson}, \&
  {Lin}}]{Terquem(1998)}
{Terquem}, C., {Papaloizou}, J.~C.~B., {Nelson}, R.~P., \& {Lin}, D.~N.~C.
  1998, \apj, 502, 788

\bibitem[{{Tokovinin} \& {Moe}(2020)}]{TokovininMoe(2020)}
{Tokovinin}, A., \& {Moe}, M. 2020, \mnras, 491, 5158

\bibitem[{{Torres} {et~al.}(2010){Torres}, {Andersen}, \&
  {Gim{\'e}nez}}]{Torres(2010)}
{Torres}, G., {Andersen}, J., \& {Gim{\'e}nez}, A. 2010, \aapr, 18, 67

\bibitem[{{Triaud} {et~al.}(2017){Triaud}, {Martin}, {S{\'e}gransan},
  {Smalley}, {Maxted}, {Anderson}, {Bouchy}, {Collier Cameron}, {Faedi},
  {G{\'o}mez Maqueo Chew}, {Hebb}, {Hellier}, {Marmier}, {Pepe}, {Pollacco},
  {Queloz}, {Udry}, \& {West}}]{Triaud(2017)}
{Triaud}, A. H.~M.~J., {Martin}, D.~V., {S{\'e}gransan}, D., {et~al.} 2017,
  \aap, 608, A129

\bibitem[{{Van Eylen} {et~al.}(2016){Van Eylen}, {Winn}, \&
  {Albrecht}}]{VanEylen(2016)}
{Van Eylen}, V., {Winn}, J.~N., \& {Albrecht}, S. 2016, \apj, 824, 15

\bibitem[{{Verbunt} \& {Phinney}(1995)}]{VerbuntPhinney}
{Verbunt}, F., \& {Phinney}, E.~S. 1995, \aap, 296, 709

\bibitem[{{Vidal} \& {Barker}(2020{\natexlab{a}})}]{VidalBarker(2020a)}
{Vidal}, J., \& {Barker}, A.~J. 2020{\natexlab{a}}, \mnras, 497, 4472

\bibitem[{{Vidal} \& {Barker}(2020{\natexlab{b}})}]{VidalBarker(2020b)}
---. 2020{\natexlab{b}}, \apjl, 888, L31

\bibitem[{{Windemuth} {et~al.}(2019){Windemuth}, {Agol}, {Ali}, \&
  {Kiefer}}]{Windemuth(2019)}
{Windemuth}, D., {Agol}, E., {Ali}, A., \& {Kiefer}, F. 2019, \mnras, 489, 1644

\bibitem[{{Winn}(2010)}]{Winn(2010)}
{Winn}, J.~N. 2010, {Exoplanet Transits and Occultations}, ed. S.~{Seager}
  ({University of Arizona Press}), 55--77

\bibitem[{{Witte} \& {Savonije}(1999)}]{WitteSavonije(1999b)}
{Witte}, M.~G., \& {Savonije}, G.~J. 1999, \aap, 350, 129

\bibitem[{{Witte} \& {Savonije}(2001)}]{WitteSavonije(2001)}
---. 2001, \aap, 366, 840

\bibitem[{{Witte} \& {Savonije}(2002)}]{WitteSavonije(2002)}
---. 2002, \aap, 386, 222

\bibitem[{{Wu}(2005)}]{Wu(2005b)}
{Wu}, Y. 2005, \apj, 635, 688

\bibitem[{{Wyrzykowski} {et~al.}(2003){Wyrzykowski}, {Udalski}, {Kubiak},
  {Szymanski}, {Zebrun}, {Soszynski}, {Wozniak}, {Pietrzynski}, \&
  {Szewczyk}}]{ogle-lmc}
{Wyrzykowski}, L., {Udalski}, A., {Kubiak}, M., {et~al.} 2003, \actaa, 53, 1

\bibitem[{{Zahn}(1966)}]{Zahn(1966)}
{Zahn}, J.~P. 1966, Annales d'Astrophysique, 29, 489

\bibitem[{{Zahn}(1975)}]{Zahn(1975)}
---. 1975, \aap, 41, 329

\bibitem[{{Zahn}(1977)}]{Zahn(1977)}
---. 1977, \aap, 500, 121

\bibitem[{{Zahn}(1989)}]{Zahn(1989)}
---. 1989, \aap, 220, 112

\bibitem[{{Zahn} \& {Bouchet}(1989)}]{ZahnBouchet(1989)}
{Zahn}, J.~P., \& {Bouchet}, L. 1989, \aap, 223, 112

\bibitem[{{Zanazzi} \& {Wu}(2021)}]{ZanazziWu(2020)}
{Zanazzi}, J.~J., \& {Wu}, Y. 2021, \aj, 161, 263

\end{thebibliography}

\end{document}